\documentclass[a4paper,11pt]{article}
\pdfoutput=1
\usepackage{amsmath,amsthm,latexsym,amssymb,amsfonts,epsfig}
\usepackage{fontenc,dsfont,layout}
\usepackage{hyperref}
\usepackage{psfrag}
\usepackage[cp1250]{inputenc}
\usepackage{graphicx}
\usepackage{color}
\usepackage{ulem}

\numberwithin{equation}{section}

\newcommand{\red}{\textcolor{\red}} 

\author{Anna Kornakiewicz${}^1$, Jakub Mielczarek${}^2$, Adam Zadro{\.z}ny${}^{3,4}$ \\
${}^1$Postgraduate School of Molecular Medicine, Warsaw Medical University, \\
{\.Z}wirki i Wigury 61, 02-091 Warsaw, Poland \\
${}^2$Institute of Physics, Jagiellonian University, {\L}ojasiewicza 11, 30-348 Cracow, Poland \\
${}^3$Center for Gravitational Wave Astronomy, UTRGV, Brownsville 78-520 TX, USA \\
${}^4$National Centre for Nuclear Studies, Andrzeja So{\l}tana 7, 00-581 Otwock-{\'S}wierk, Poland
}
\title{A concept of biopharmaceutical nanosatellite}

\begin{document}

\maketitle

\begin{abstract}
The article is a short overview of a proposal of a CubeSat type nanosatellite
designed to conduct biopharmaceutical tests on the low earth orbit. Motivations
behind the emerging demand for such solution nowadays and in the close
future are emphasized. The possible objectives and challenges to be addressed
in the planned biopharmaceutical CubeSat missions are discussed. In particular,
it is hard to imagine progress of the space tourism and colonization of Mars
without a wide-ranging development of pharmaceutics dedicated to be used in space.
Finally, an exemplary layout of a 3U type CubeSat is presented. We stress that, thanks
to recent development in both nanosatellite technologies and lab-on-a-chip
type biofluidic systems the proposed idea becomes now both feasible and
relatively affordable.

\end{abstract}

\section{Introduction}

A huge progress has been made over recent years in the development of
various functional tissues and organoids which may play significant role in
the pre-clinical stage of drug development and personalized medicine. The
progress in the discipline benefited significantly from development of such
new technologies as 3D bioprinting \cite{Murphy}, microfluidics \cite{Dittrich} and
\textit{organ-on-a-chip} \cite{Bhatia}. The technologies became a basis of viable
business models which prognose further rapid development of the commercial
applications in various branches of medicine, pharmacy and biotechnology.

Another unprecedented business activity is currently observed in the domain
of space technologies. This is mostly stimulated by development of information
accumulating, transmitting and processing solutions employing the fully
functional nanosatellites, in particular those in the CubeSat standard. This
includes swarm CubeSats systems devoted to Earth imaging and global
internet access. Furthermore, the cargo delivery to the Low Earth Orbit (LEO) is
taken over by such companies as Space X. Further solutions are in
advanced stage of development by such companies as Rocket Lab,
Blue Origin, Virgin Galactic or Stratolaunch Systems. These new possibilities
will not only significantly decrease costs of placing nanosatelites on Earth orbit
but also contribute to development of human space activity including space
tourism.

But there is even more about space going on nowadays. The plans of building
settlements on Moon and Mars are becoming more serious, clear and feasible
\cite{Ehrenfreund}. This is mostly because of the the effort of ESA, CNSA,
NASA and Space X.

Therefore, in the eve of space colonization \cite{Musk}, we have to ask ourselves
if the humans are ready for this and if not what kind of the difficulties have to
overcome by addressing them early enough to be ready on time. In our opinion,
one of such a challenge is development of different kind of drugs which will be
used during the manned long term space activity, including colonization of Moon
and Mars as well human activities related with the space mining in the deep space.
This is motivated by the following main facts:
\begin{itemize}
\item During the long-drawn space missions humans are exposed to elevated
cosmic radiation level, which increases possibility of tumor development.
\item The microgravity conditions affect significantly the metabolic processes
which may result in diseases and injuries.
\item Very limited research has been conducted to determine whether the expected
Earth-based pharmacokinetics and pharmacodynamics of a drug are altered in a
microgravity environment.
\item Action of the pharmaceutics is affected by the cosmic stresses such as microgravity
and cosmic radiation.
\item There is no data currently to provide a basis for clinical recommendations.
\end{itemize}

Therefore, there is a need to develop a set of drugs which are designed to be used in
the space environment. Undertaking the challenge may also advance use of drugs on
the Earth. Testing platform for drugs in space is to be build. As we are going to discuss
below, such objective seem to be possible to achieve thanks to the combination of
nanosatellite technologies with the 3D cell cultures designed to the drug testing applications.

\section{Objectives and challenges}

The idea of using CubeSats to conduct relatively cheap astrobiological research in space
is not new. For instance, such missions as GenSat-1 \cite{GenSat1}, PharmaSat 1
\cite{PharmaSat1} and O/OREOS \cite{Oreos} were conducted successfully in the recent
years. Further CubeSat missions as EcAMSat \cite{EcAMSat} and BioSentinel \cite{BioSentinel}
are in the advanced stage of preparation. The BioSentinel experiment is especially interesting
since it is going to be carried beyond LEO and launched by the first flight
of Space Launch System (SLS) rocket \cite{SLS}, which is designed to conduct both Lunar
and Martian missions. The BioSentinel is a 6U type mission which contains a microfluidic card
equipped with 3-color LED detection system allowing to analyze metabolic properties of the
yeast \textit{S. cerevisiae} in the deep space. The reference system will be placed at the
International Space Station (ISS), where the microgravity is similar but the amount of radiation
is suppressed with respect to the deep space environment due to the effect of Earth magnetic field.

Investigation of the effect of cosmic radiation on basic model with yeast might be
the first step towards testing drugs alone and in combination. There is no basic data in
literature on effect of radiation on DNA structure in cells. Radioprotective mechanism
of action of drugs on DNA structure level is to elucidate and yeast model may be suitable \cite{bib12}.
When basic information on  genetic alteration under radiation condition is gathered,
next step will be to test effect of drugs in  3D  cell culture which mimics natural
development of tumor. Currently, the most broadly study groups of drugs in oncology are
immunologic and antiangiogenic drugs \cite{bib13}. Planing to test novel anti-tumor drugs which
exhibits immunomodulatory and anti-angiogenic properties, challenge is to build precise
model of tumor environment in cell culture, therefore cell culture should not consist
only of cancer cells, but also epithelial cells of vessels and cells of immunologic system.
Also extremely interesting seems to be gaining insight into cancer stem cell biology
in space condition, therefore next dimension of cellular model is to incorporate
cancer stem cells.

Basic test to perform on cell culture is alamarBlue\circledR. This is cell assay which
allows quantitatively measure of cell viability \cite{bib14}. Changes in viability are to detect by
absorbance plate reader. Also genetic expression of cells may be measured with other
types of essays.

The novelty we are proposing to implement in the space environment are the cell
cultures equipped with (micro-)fluidic system, designed to conduct biopharmaceutical
studies.  While prokaryotic cells such bacteria (e.g. \textit{E. coli}) and fungi
(e.g. yeast \textit{S. cerevisiae}) provide a well established standard for studying
impact of radiation on genetic information (e.g. double strand breaks (DSB)), testing of
 pharmaceutics require to use more sophisticated eukaryotic cell cultures.

The problem is, however, that conducting animal cell cultures, even in laboratory
environment requires special conditions and processing. Miniaturization of such
laboratory setup to the side of, let say, 2U and choice of adequate cell cultures
is, therefore, a serious challenge.

The type of cell cultures which  are from one side the most stable ones and at the
same time also one of the most relevant to be studied in the space environment
and cancer cells. One of the characteristic features of the tumor cells is their immortality
which means that they divide continuously.  In spite of this properties culturing cancer
cells is demanding, even in laboratory cultures - they require special condition and
treatment to growth. Maintaining continuity of cell culture in space is much more
demanding, as it needs automation in changing media and passaging. This is to
ensure by culturing in specialized fluidic systems. However, the recent advances in
development of the \textit{cancer-on-a-chip} \cite{Zhang,Tsai} systems give us the
basics to claim that the required miniaturization of the biopharmaceutical device is
possible to achieve.

In order to give an intuition of some relevant orders of magnitude, let us consider a
cell culture confined to the volume of 1 cm$^3$ which is going to be conducted for
the period of up to one month. This period should be long enough to already see
some effects of microgravity and radiation. Typically, around 1000 cancer cells is
plated in let us say 4 ml of medium and there cells can divide in logarithmic rate at
the beginning with plateau growth later occurring depending on refeeding 
(the process can be modeled by the Gompertz curve). For the period of week number 
of cells can increase 10 times, in the first week growth is faster - in the second week 
growth rate can be 2 times lower. Usually, once a week around 3-4 day medium 
should be changed - than rate of cell growth slightly increase. Therefore, if we 
want to simulate laboratory conditions in nanosatellite, we should ensure medium 
is changed for times in the period of month. Culturing media in suspension should 
ensure 90\% viability rate in conditions 37 deg C, 5\% CO$_2$, 16-19\% O$_2$. 
Major importance in the project is development of suitable suspension-culture media 
that requires little or no additional task to maintain culture.

Detailed analysis of dynamics of the cancer growth as function of control 
parameters has to be performed in order to optimize the experimental setup. 
The preliminary steps can be performed with the use of the Gompertz of the 
tumor growth in constrained environment. In such a case the number of cancer 
cells in time is given by the following formula \cite{Yorke,Enderling}:
\begin{equation}
N(t) =N_0 \exp \left\{ \ln \left(\frac{N_{\infty}}{N_0}\right) \left[1-\exp (-\lambda t) \right] \right\},
\label{Gompertz}
\end{equation}
where $N_0$ is the initial number of cancer cells and $N_{\infty}$ is the upper
bound on the number of cells in the culture. In particular, in the considered 
exemplary setup with the maximal volume constrained to 1 cm$^3$ one can 
estimate that $N_{\infty} \sim 10^9$. Extracting from (\ref{Gompertz}) the initial 
doubling time 
\begin{equation}
{\rm DT} := - \frac{1}{\lambda} \ln\left [ 1 -\frac{\ln 2 }{\ln \left(\frac{N_{\infty}}{N_0}\right)}  \right], 
\end{equation}
and comparing it with doubling time scales for cells culture under consideration 
(which may range from hours to hundreds of days) the parameter $\lambda$ of the 
model can be determined. Based on this the experimental details can be fixed 
such that the plateau of the culture size is reached in the period of around one 
month. Then, impact of the anti-tumor treatment on dynamics of the tumor growth
can be modeled by generalizations of Eq. \ref{Gompertz} (see e.g. \cite{Enderling}). 

\section{CubeSat overview}

\begin{figure}[ht!]
\centering
\includegraphics[width=16cm,angle=0]{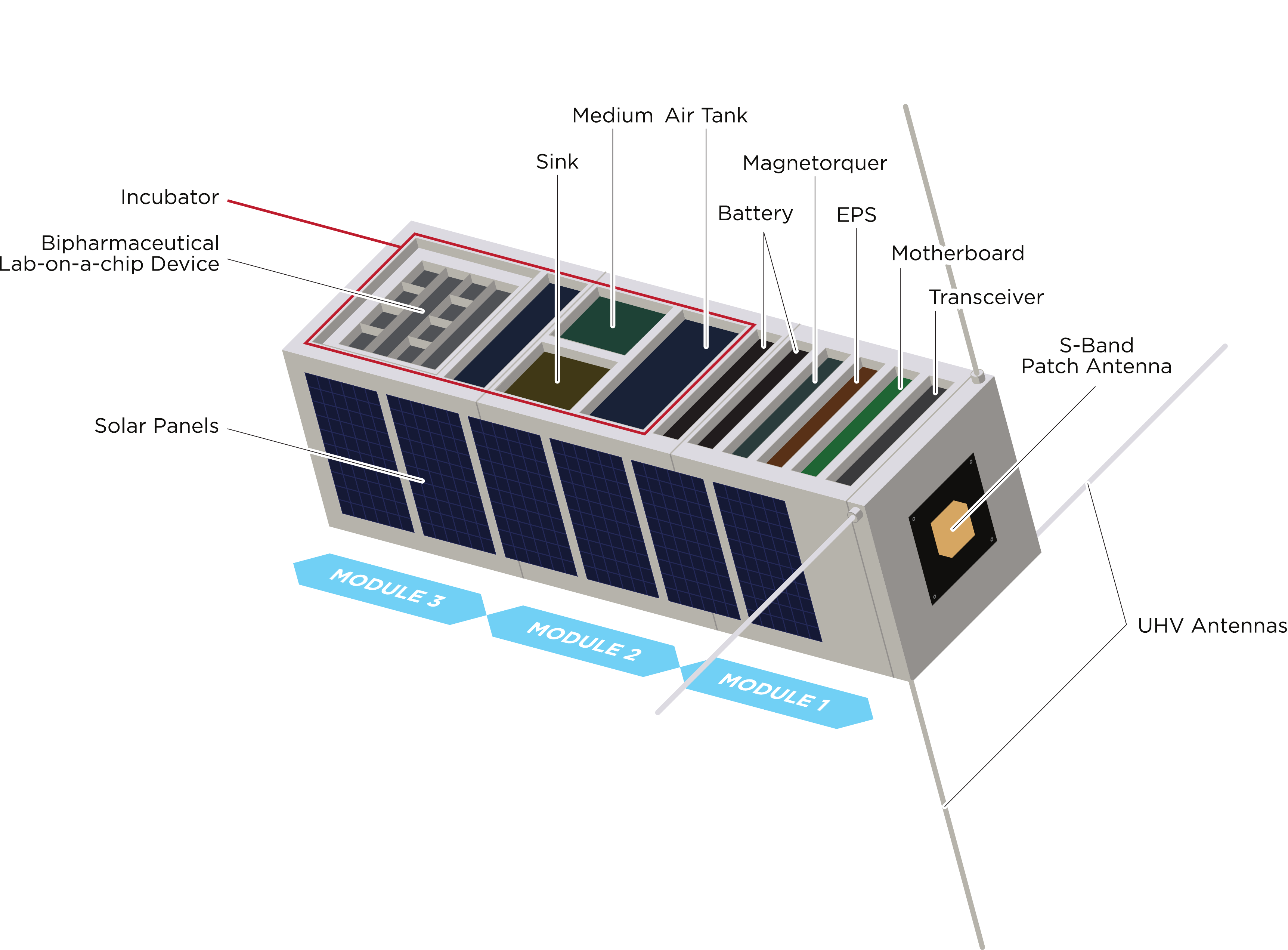}
\caption{Block structure of the 3U biopharmaceutical CubeSat. Each module is of the 1U size.}
\label{Scheme}
\end{figure}

According to our estimates, the 3U size in the CubeSat standard is sufficient to
accommodate all key systems of the basic biopharmaceutical nanosatellite.
An exemplary CubeSat contains three functional modules, each of the 1U size
(10$\times$10$\times$10  cm$^3$), as presented in Fig. \ref{Scheme}

The Module 1 contains system responsible for: control, data processing, communication,
orientation and energy supply.  Communication with the satellite can be conducted
simultaneously in the UHF (or/and VHF) band and the S-band. While the basic control
uses the UHF in downlink and uplink mode the S-band in the downlink mode is devoted
to transmit scientific data. The S-band transmission is provided by the patch antenna
working at the conventional 2.4 GHz WiFi frequency. There is a few (about 5) observational
5 min windows a day. The 9600 bps downlink transfer of information is planned. This
means that, within a single observational window around 350 kB of data can be
transferred, including control information and error correction. This amount of information
is expected to be sufficient to cover partially pre-processed data from the scientific
instruments. The transceiver module is responsible for managing transmission at both
UHF and S-band  frequencies. The data are managed and processed at the
Motherbord connected via I$^2$C port with the other subsystems.  Among them, the
Magnetorquer will allow to fix correct orientation (towards to Earth) of the S-band
antenna. Furthermore, the Module 1 contains Electrical Power System (EPS) and 
batteries.

The Module 2 contains: reservoir of cell culture medium, air (O$_2$ and CO$_2$
mixture) tank, sink for the fluidic system and temperature stabilization system.
The estimated volume of the cell culture medium reservoir needed to conduct
one month long studies is 100 ml.

The Module 3 contains: experimental unit equipped with fluidic system and tumor
cell culture, camera for tumor growth imaging, LED absorbance detectors for metabolism
monitoring, dose of the pharmacological active substance. Details of the experimental
setup are to be specified in the forthcoming steps of the development of the project.

Both experimental elements contained in Module 1 and Module 2 are enclosed in
temperature stabilization system providing incubating conditions for the cell culture, 
i.e. 37 deg C. The system is crucial since it must provide stable conditions for 
the whole experimental setup. The temperature stabilization system will be also the 
most energy consuming.  However, due to other constraint, the system has to be 
designer such that its consumption does not exceed around 2W. Achieving such
a goal will require application of specialized insulating solutions to prevent against 
both cooling and excessive heating of the system. 

All four longer sides of the CubeSat are covered by photovoltaic cells providing in total
around 7 W with 3V supply voltage.  Furthermore, at least 25 Wh in the battery pack
will be needed.

The proposed construction is designed to operate at the LEO for a period up to
around 3 months. Depending on the orbit details, there might be need to complement
the nanosatellite with an additional deorbitation system.  Furthermore, one has to
keep in mind that while the LEO provides higher amount of radiation comparing to
the Earth, testing the effects related e.g. with human exploration of Mars requires to
go beyond LEO, where the amount or radiation is higher. This will be, in particular,
the scope of the 6U type BioSentinel experiment.

\section{Summary}

The purpose of this article was to emphasis an emerging need to start performing
biopharmaceutical tests in space. We have stressed that combination of two recently
rapidly expanding technologies i.e. lab/organ-on-a-chip and nanosatellites may make
such studies accessible and affordable. The expected cost of a single mission is
to be not greater that 2 million Euro. Such relatively low costs space experiments if
well designed can provide scientifically unique and practical data with a potential
for successful commercialization. In particular, it is hard to imagine progress of the
space tourism and colonization of Mars without a wide-ranging development of
pharmaceutics dedicated to be used in space. In particular, due to the increased risk
of cancer due to the cosmic radiation, anti-tumor drugs are to be developed.


\begin{thebibliography}{99}

\bibitem{Murphy}
S.~V.~Murphy, A.~Atala, ``3D bioprinting of tissues and organs,'' Nature Biotechnology {\bf 32}, 773-785 (2014).

\bibitem{Dittrich}
P.~S.~Dittrich, A.~Manz, ``Lab-on-a-chip: microfluidics in drug discovery,'' Nature Reviews Drug Discovery volume {\bf 5}, 210-218 (2006).

\bibitem{Bhatia}
S.~N.~Bhatia, D.~E.~Ingber, ``Microfluidic organs-on-chips,'' Nature Biotechnology {\bf 32}, 760-772 (2014)

\bibitem{Ehrenfreund}
P.~Ehrenfreund et al., ``Toward a global space exploration program: A stepping stone approach,"   Advances in Space Research. Volume {\bf 49}, Issue 1, 2-48 (2012).

\bibitem{Musk}
E.~Musk, ``Making Humans a Multi-Planetary Species,'' New Space. June 2017, {\bf 5}(2): 46-61.

\bibitem{GenSat1}
C.~Kitts et al., ``The GeneSat-1 Microsatellite Mission: A Challenge in Small Satellite Design,'' 20th Annual AIAA/USU Conference on Small Satellites (2006).

\bibitem{PharmaSat1}
C.~Kitts et al., ``Initial Flight Results from the PharmaSat Biological Microsatellite Mission,'' 23th Annual AIAA/USU Conference on Small Satellites (2009).

\bibitem{Oreos}
P.~Ehrenfreund et al., ``The O/OREOS mission-Astrobiology in low Earth orbit,''  Acta Astronautica {\bf 93} (2014) 501-508.

\bibitem{EcAMSat}
https://www.nasa.gov/centers/ames/engineering/projects/ecamsat

\bibitem{BioSentinel}
https://www.nasa.gov/centers/ames/engineering/projects/biosentinel.html

\bibitem{SLS}
https://www.nasa.gov/exploration/systems/sls/index.html

\bibitem{bib12}
P.~Nemavarkar, B.~K.~Chourasia, and K.~Pasupathy, "Evaluation of radioprotective action of compounds using Saccharomyces cerevisiae," J. Environ. Pathol. Toxicol. Oncol. Off. Organ Int. Soc. Environ. Toxicol. Cancer, 
vol. {\bf 23}, no. 2, pp. 145-151, 2004.

\bibitem{bib13}
N.~S.~Vasudev and A.~R.~Reynolds, "Anti-angiogenic therapy for cancer: current progress, unresolved questions and future directions," Angiogenesis, vol. {\bf 17}, no. 3, pp. 471-494, 2014.

\bibitem{bib14}
S.~N.~Rampersad, "Multiple Applications of Alamar Blue as an Indicator of Metabolic Function and Cellular Health in Cell Viability Bioassays," Sensors, vol. {\bf 12}, no. 9, pp. 12347-12360, Sep. 2012.

\bibitem{Zhang}
Y.~S.~Zhang, Yi-N.~Zhang, W.~Zhang, ``Cancer-on-a-chip systems at the frontier of nanomedicine,''
Drug Discovery Today,  Volume {\bf 22}, Issue 9, 1392-1399 (2017).

\bibitem{Tsai}
Hsieh-Fu Tsai, et al.,``Tumour-on-a-chip: microfluidic models of tumour morphology, growth and microenvironment,''
J R Soc Interface.  {\bf 14}, 131 (2017).

\bibitem{Yorke}
E.~D.~Yorke, et al., ``Modeling the development of metastases from primary and locally recurrent tumors: comparison with a clinical data base for prostatic cancer,'' Cancer Res. {\bf 53} (13):2987-93 (1993).

\bibitem{Enderling}
H.~Enderling, M.~A.~J.~Chaplain, ``Mathematical modeling of tumor growth and treatment,'' Curr Pharm Des. {\bf 20} (30):4934-40 (2014).

\end{thebibliography}
\end{document}